\documentclass[preprint2]{aastex}
\slugcomment{submitted to {\it The Astronomical Journal}}
\begin{document}
\title{$uvbyCa$H$\beta$ CCD Photometry of Clusters. III. The Most Metal-Rich
Open Cluster, NGC 6253}
\author{Bruce A. Twarog\altaffilmark{1}, Barbara J. 
Anthony-Twarog\altaffilmark{1},
and Nathan De Lee\altaffilmark{2}}
\affil{Department of Physics and Astronomy, University of Kansas, 
Lawrence, KS 66045-7582}
\affil{Electronic mail: twarog@ukans.edu,bjat@ku.edu,ndelee@msu.edu}
\altaffiltext{1}{Visiting Astronomer, Cerro Tololo Interamerican Observatory.
CTIO is operated by AURA, Inc.\ under contract to the National Science
Foundation.}
\altaffiltext{2}{Current address: Department of Physics and Astronomy, Michigan State
University, East Lansing, MI 48824}
\begin{abstract}
CCD photometry on the intermediate-band $uvbyCa$H$\beta$ system is presented
for the old open cluster, NGC 6253. Despite a high level of field star
contamination due to its location toward the galactic center, combination of the
data from the multiple color indices with the core cluster sample derived
from radial star counts leads to the identification of a set of
highly probable, single cluster members. Photometric analysis of a
select sample of 71 turnoff stars produces a reddening value of
$E(b-y) = 0.190 \pm 0.002$ (s.e.m.) or $E(B-V)$ = 0.260 $\pm 0.003$ (s.e.m.)
from 71 stars.  The metallicity indices, $\delta$$m_1$ and $\delta$$hk$,
both identify this cluster as the most metal-rich object studied on either
system to date. Simple extrapolation of the available metallicity 
calibrations leads to [Fe/H] ranging from +0.7 to +0.9. Metal-rich isochrones
with overshoot imply an age between 2.5 and 3.5 Gyr, with an apparent
distance modulus between $(m-M)=$ 11.6 and 12.2, depending upon the isochrones 
used.  The improvement in the fit using $\alpha$-enhanced isochrones may 
indicate that the cluster [Fe/H] is closer to +0.4, but the photometric indices 
are distorted by an elemental distribution other than a scaled solar. The 
galactocentric position of the cluster, in conjunction with data for
other clusters and Cepheids, is consistent with the inner disk reaching
and maintaining a metallicity well above solar since the early history of
the disk, unlike the solar neighborhood.
\end{abstract}

\section{INTRODUCTION}
The shared properties of distance, age, and composition for stars within
a cluster supply justification for their study within
the context of stellar and galactic
evolution. An example of how a well-defined and homogeneously-analyzed
cluster sample can shed light on galactic evolution is provided by
\citet{TAT97}, wherein analysis of over 70 open clusters demonstrates the
existence of structure within the galactic abundance gradient, structure
that has been corroborated most recently through the use of Cepheids by
\citet{AN02}.
Despite the unusual combination of high precision and sample size for
the collected data, critical regions of the galactic disk remain
inadequately sampled. Moreover, only a modest fraction of the open
clusters within observational reach have their fundamental
parameters determined to a degree of reliability that can satisfy the needs of
Galactic studies tied to the global characteristics of the
population. 

Though the number of broad-band photometric surveys of individual
clusters with color-magnitude diagrams (CMD) of sufficient precision
has increased significantly in the last five years with the expanded 
availability of larger-format CCD's, the interpretation of these data
has been hampered by the absence of a comparable progress in determinations
of the critically important properties of reddening and metallicity,
linchpins to age and distance. Attempts to constrain the parameters
using CMD morphology and comparison to theoretically synthesized CMD's
have had some success \citep{TO91,BO90,GO96,MA97}, though the answers drawn 
from different models do
not always concur and the uncertainties in the parameters remain larger
than desirable for many projects.

To improve this situation for individual clusters and to expand and test the
sample delineating the galactic structure outlined by \citet{TAT97}, a program
of CCD photometry of open clusters on the extended Str\"{o}mgren 
intermediate-band
($uvbyCa$) and H$\beta$ systems has been undertaken. Examples of the initial
work illustrating the value and precision attainable with careful analysis
have been given for the 
metal-rich open cluster, IC 4651 \citep[hereinafter referred to
as Paper I]{AT00a} and the globular cluster, NGC 6397 
\citep[hereinafter referred to as Paper II]{AT00b}. These clusters 
were selected to sample the 
extreme range of cluster
parameters with objects that had been reasonably well-studied by other
photometric and spectroscopic means. Our first program cluster with
no intermediate-band photometric work or reliable high-resolution
spectroscopic abundances is the distant open cluster, NGC 6253.

NGC 6253 has been the focus of three broad-band photometric surveys
to date, \citet{BR97}, \citet{PI98}, and \citet{SA01} (hereinafter
referred to as BR, PI, and SA, respectively). It is of interest
because, as carefully demonstrated in BR, the cluster
is located toward the galactic center ($l^{II} = 336\arcdeg$, $b^{II} 
= -6\arcdeg$),
significantly older than the Hyades, and super-solar in metallicity, while
exhibiting a richly-populated turnoff and giant branch.
Despite the concurrence of opinion regarding the broad features of the
cluster, examination of previous work reveals a wide range for 
the key parameters of interest. 
Previously published work produces plausible reddening 
estimates ranging from $E(B-V)$ = 0.20
to 0.32, metallicity from solar abundance to almost triple the solar 
value, and an age from 2 to 5 Gyr; we will discuss the specifics of 
each property and its past derivation in Sec. 4.
Since traditional linear abundance gradients imply a change in
the mean [Fe/H] of less than 0.3 dex over a 4 kpc galactocentric 
range, definitive evaluation of any structure within 2 kpc of the 
sun requires [Fe/H] data of significantly higher accuracy than is currently
available for NGC 6253, for a large number of clusters. Moreover,
a definitive metallicity is important in defining the value of the
cluster as a template for younger, metal-rich, extragalactic populations
and as a potential source of stars with planetary systems.  

The underlying basis for the use of intermediate-band photometry is simple.  The brighter observational limit imposed by the narrower bandpasses is, in part, compensated by the higher precision with which one may disentangle the stellar parameters. DDO photometry \citep{TAT97} recommends itself because it focuses on the brightest stars within a cluster, the red giants, and the calibration of the system is exquisitely tied to the spectroscopic abundances of the extensive sample of field giants near the sun \citep{TAT96}. The weaknesses of the system are the small sample of giants often present in many poorly populated clusters observed within a field rich with cluster non-members and the possibility that post-main-sequence evolution may alter the atmospheric abundances of some stars. 

The extended Str\"{o}mgren and H$\beta$ systems exploit the rich 
population of main sequence
stars and have been well-calibrated for spectral types ranging from B through
G \citep{CR75,CR78,CR79,OLS88,SN89}, the temperature range 
typified by the turnoff in most open clusters. No open cluster has 
been identified that exhibits a range in [Fe/H]
or a difference in [Fe/H] between the turnoff stars and the red giants when
the abundances are based primarily upon the Fe lines. The larger population
of main sequence stars can lead to easier identification of a probable
sample of cluster members while reducing the standard error of the mean for
the final cluster parameters. The multiple indices also enhance the elimination
of probable non-members from the final sample. For an example of how
successful this approach can be for a typical open cluster, the reader is
referred to Paper I. 

Section 2 contains the details of the CCD observations and their
reduction and transformation to the standard system. In
Sec. 3 we discuss the CMD and begin the process of identifying the
sample of probable cluster members. Sec. 4 contains the derivation of the
fundamental cluster parameters of reddening and metallicity. In
Sec. 5, these are combined with broad-band data to derive the distance and
age through comparisons with theoretical isochrones. Sec. 6 summarizes 
our conclusions in the context of
our current understanding of the evolution of the galactic abundance 
gradient. Preliminary results for NGC 6253 presented in \citet{TAD}
are superseded by this discussion. 

\section{The Data}

\subsection{Observations}
The new photometric data for NGC 6253 were obtained using the Cassegrain-focus 
CCD imager on the National Optical Astronomy Observatory's 0.9-m telescope
at Cerro Tololo Interamerican Observatory. We used a 
Tektronix 2048 by 2048 detector at the $f/13.5$ focus
of the telescope, modulated by CTIO's $4'' \times 4''$ $uvby$ filters 
and our own $3'' \times 3''$ H$\beta$ and $Ca$ filters.  The field size is 
$13.5'$ on a side.  Frames were obtained in April 1999 with all filters 
except the H$\beta$ filters; frames obtained in May 2000 included all 
bandpasses except $Ca$.  In all, 11 to 22 frames were acquired of NGC 6253
for each filter with the following total exposure times:  37, 63, 71 and 159 
minutes for $y, b, v$ and $u$, 83 minutes for $Ca$, and 40 and 183 minutes
for the H$\beta$ wide and narrow filters.

Standard IRAF routines were used to perform initial processing at the 
telescope; a fairly comprehensive description of these steps and the procedures
by which we merge photometry based on profile-fitting algorithms may be found
in Paper I.   

\subsection{Reduction and Transformation}
The procedures outlined in Paper I yield photometric 
indices of very high internal
precision.  We outline here the steps taken to transform these instrumental
indices to the standard systems, and will separately note specific issues 
pertaining to
the reduction of $uvby$ indices, $hk$ indices, and H$\beta$ indices following
this overview.

In addition to our program fields, we observe standard stars selected from
the photometry catalogs of \citet{OLS83,OLS93,OLS94} and \citet{TAT95} 
and in open clusters on each photometric night. These are
used to determine extinction solutions for each photometric night 
for all indices other than H$\beta$. Aperture photometry is obtained for
every standard star and program
cluster on all but the deepest exposure frames using aperture
radii ranging from 8 to 13 pixels centered on the star, along with sky
measurements using annuli enclosing a comparable area. 
Following correction for atmospheric extinction and the
exclusion of stars with the most serious crowding, average magnitudes in each
bandpass are constructed for all stars.  From the mean magnitudes, 
mean indices are constructed for stars in every field on each night.
For a program cluster observed on a photometric night, the mean difference 
between the aperture photometry and the profile-fit-based photometry is 
then determined 
for each index based on a selected sample of relatively uncrowded stars.

In determining the transformation equations between the instrumental
and standard system, our preferred approach is to use the
photometry within the open clusters to establish the slopes of 
the transformation relations and, if
possible, appeal to photometric catalog values of single-star standards 
to establish the zero point of the calibration equations.  With 
these transformation 
equations, indices constructed from all the frames may be 
converted to standard values.  By inference,
the profile-fit photometry may be calibrated with the same relations.

H$\beta$ frames of NGC 6253, as well as several standard fields 
in open clusters,
were obtained on the night of May 2, 2000.  The transformation 
equation for conversion of
aperture-based H$\beta$ indices is defined collectively 
by 74 stars in NGC 6231 \citep{SH83},
NGC 6664 \citep{SC82}, M67 \citep{NT87} and NGC 6633 \citep{SC76}. 
The standard H$\beta$ indices range in
value from 2.53 to 2.87, leading to an equation of the form 
$\beta = 1.157 \beta_i$, with a dispersion of only $\pm$0.015 mag about 
the mean line.  
The mean difference between aperture-based H$\beta$ indices from this night
and the H$\beta$ indices in the profile-fit-based photometry set of NGC 6253 was
determined with a standard error of the mean $\leq 0.001$ mag.

The only $Ca$ frames of NGC 6253 taken on a photometric night were obtained
on April 22, 1999.  Transformation of the $hk$ indices is tied to the 
observations of field stars found in the catalog of \citet{TAT95}.  
Data from several nights in April 1999 and May 2000 were 
combined and analyzed, leading to a transformation relation of the form
$hk = 1.05 hk_i$. Four field star standards observed
on April 22, 1999 established the zero point for the equation calibrating the
$hk$ indices specifically for that night, with a standard error in 
the mean for the intercept of $\pm$0.012. The mean difference 
between the aperture-defined $hk$ indices and the profile-fit-based $hk$ indices 
was derived with a standard error in the mean value of only $\pm$0.005.
Comparable differences between aperture photometry and profile-fit-based
values for other Str\"omgren indices were determined with precisions
indicated by standard errors of the mean for $V$ and $b-y$ of 0.002 mag, and
0.003 mag for $m_1$ and $c_1$.
 
Calibration of $b-y$, $m_1$ and $c_1$ indices is generally challenging, in part
because the parameter space of color, abundance, and luminosity class 
must be delicately covered by a wide range of standards.  Following normal 
practice, we break the calibration equation solutions at a color 
of $(b-y)_0 = 0.42$. While past experience has shown that cooler 
dwarfs generally have 
a slope for the $b-y$ transformation relation that is
shallower than that for the warmer dwarfs and cool giants, no 
cool dwarf members of NGC 6253 are expected to be within our survey sample.
The transformation equation for warmer dwarfs was applied to
NGC 6253 stars with $b-y$ less than 0.62, approximately the
intrinsic color break of 0.42 with a reddening adjustment of 0.20.
Transformation equations for $m_1$ and $c_1$ indices for redder stars,
presumed to be giants, were determined separately based on field star
standards from the \citet{OLS93} catalog.

Several open clusters observed on May 4 and 5, 2000 were used to delineate
the form of these calibration equations.  Specific references are
NGC 6231 \citep{SH83}, NGC 6633 \citep{SC76}, M67 \citep{NT87} and
NGC 3680 \citep{NI88}. In addition, observations on April 22, 1999 in
IC 4651 were used to calibrate NGC 6253 indices obtained from
aperture photometry on that night by comparison to values published in
Paper I.  NGC 6253 was not observed on May 5, 2000.
Field star standards from the catalogs of \citet{OLS83} and \citet{OLS93}
were observed on both nights in May 2000 and form the basis of the 
intercepts for the calibration equations.  

It became apparent that the conditions for part of the night 4 May 2000
must have deteriorated from ideally photometric.  For example, the standard 
deviations for the calibration equation zero point values, based on nine 
field star standards observed on that night, were 
distressingly large, 0.04 at best
and as high as 0.07 for $m_1$. We constructed an independent check of
the calibration of Str\"{o}mgren indices based 
on the April 22, 1999 photometry in IC 4651 and several field star standards.
We were able to estimate corrections to the zero points for the
transformation equations with precisions indicated by the following 
standard errors of the mean for the zero points: $\pm$0.002 for $V$ 
and $b-y$, and $\pm$0.004 for $m_1$ and $c_1$.

Final photometric values for the stars brighter than $V \simeq 18.3$ are given
in Table 1. Pixel coordinates for the stars are on the (X,Y) system for
the cluster developed by WEBDA and WEBDA identification numbers are
provided whenever available. The standard errors of the mean as a 
function of $V$ are presented in Fig. 1 for all of the 
indices. The errors are derived from the frame-to-frame dispersion for
each filter used to construct the indices rather than the formal errors
from IRAF. Clearly, the internal errors are excellent for $V \leq 17.5$ for all indices.

\subsection{Comparison to Previous Photometry}
No intermediate-band photometry on any system has been published for
NGC 6253. However, the $V$ magnitude derived from the $y$ filter
can be compared to the published broad-band data. As detailed in the
Appendix, there are systematic differences with all the published $V$
photometry, though our data are located near the approximate mean of the three
photometric surveys. All of the broad-band studies exhibit non-negligible
offsets with respect to each other and, in one case, a significant color dependence
is found among the residuals. We will return to this point in Sec. 5 when
we discuss the cluster match to the theoretical isochrones.  

\section{The Color-Magnitude Diagram: Thinning the Herd}

The CMD for all stars with $V$ brighter than $\sim 18.3$ and 
at least 2 observations each in
$b$ and $y$ is presented in Fig. 2. Filled circles are stars brighter
than $V$ = 17 that have larger than expected variation in $V$ at their
magnitude and have at least 5 observations in $y$ and multiple 
observations in all filters. The gross features of the cluster CMD are
observable and confirm the results found in the broad-band surveys. The
turnoff near $V \sim 15$ exhibits a sharp hook to the red below a well-defined
gap. The subgiant branch is well populated, sloping downward and to the red.
The red giant clump is easily identifiable near $V \simeq 12.6$, with a bright
red giant branch that may extend to $V \simeq 11.3$ before turning over.
The morphological resemblance to another definitive old open cluster,
M67, is striking.

Unfortunately, due to its galactic location, contamination
of the CMD by non-members is a dominant problem for any analysis.
Below $V$ = 16.5, the cluster main sequence merges into the amorphous
band defined by background field stars, while field subgiants and
giants dominate for $b-y$ redder than 0.7 and $V \geq 15$. The problem
is similar in nature to that faced in Paper I for IC 4651, with the
complication that the cluster turnoff lies much closer in color and
magnitude to the field star distribution due to the greater distance of
NGC 6253. 

As discussed in Paper I, the focus of our analysis of
this cluster, and all clusters in the program, is the precise
determination of the primary cluster characteristics of reddening and 
metallicity followed, when possible, by 
distance and age. Photometric derivation of
these parameters requires  color indices of high precision for single-star members
of the cluster. If we have achieved the photometric accuracy we require,
the dominant issue becomes the isolation of highly probable members of
the cluster from the field distribution. Procedures that identify and
eliminate probable non-members are acceptable even if they remove some
cluster members. Keeping non-members out is more important than keeping
members in since completeness of the sample is not a issue.
In the sections that follow, we will refine the sample in Fig. 2 to 
reach a point where the probable cluster members alone define the features of
the CMD.

\subsection{Thinning the Herd: The Radial Distribution}
In the absence of kinematic information to establish cluster membership, the
next best option is to use the radial distribution of cluster stars to select 
the area with the highest proportion of members.  Previous photometric surveys have established estimates of the cluster diameter, beginning with an estimate 
of $5'$ from \citet{LY87}.  The limiting field size of CCD's has been a factor
in subsequent photometric studies.  BR studied two overlapping fields
that were $6.8' \times 6.8'$, one centered on the cluster and one
north of center.  SA imaged a cluster field $2.1' \times 3.3'$ along
with a comparison field $10'$ to $15'$ away from the cluster center.
PI observed a field approximately $4' \times 4'$. With the exception of BR,
the fields studied have been smaller
than the proposed cluster diameter and, even in the case of BR, the counts
beyond the cluster core cover a small portion of the area around the
cluster. The location of the cluster center has been selected
by a simple visual
fit in the case of BR, while PI and SA fixed the center by adjusting
the coordinates to maximize the cluster density profile. The locations found
by BR and SA are similar, but differ from that of PI by almost $1'$.

With a larger field size, we hoped to have an easier time establishing
the profile and center of the cluster. Coordinates for stars brighter than
limits varying from $V_{lim} = 15.0$ to 18.0 were used to determine
the cluster center which would optimize the concentration of the cluster
distribution within the smallest radius.  The cluster center we found is
similar to that found by BR and SA, though the uncertainty in the exact
location seems to be larger than claimed by SA, due to the small
number of bright stars defining the center.  The cluster blends into the
background at a diameter of $7'$, larger than found by \citet{LY87}.

Due to the modest portion of the CCD field occupied
by the cluster, the dominance of the CMD by non-members is easily
understood. The cluster excess above background is even smaller 
for $V \geq 17.5$, showing an increasing deficiency of lower
mass stars compared to expectations from normal luminosity functions.
To maximize the percentage of cluster members, our first cut will remove
all stars located more than $100''$ away from the cluster center.
Simultaneously,
we will identify all stars that have standard errors in the mean for $b-y$
greater than 0.010 for later elimination. The resulting CMD is presented in
Fig. 3, where crosses identify stars that fail to meet the error cut.

The primary features of the cluster CMD are significantly enhanced, though
confusion caused by field star contamination remains an issue below $V$
= 17. Beyond the features noted earlier, one can now identify a binary
sequence located about 0.7 mag above the expected position of the main
sequence. The exact location of the main sequence is still a source of
controversy due to the band of field stars that extends from $b-y$ = 0.55
the region redward of the subgiant branch.  

\subsection{Thinning the Herd: CMD Deviants}
Given the high precision of the $b-y$ indices, if the cluster sample 
dominated that of the field stars, one could define the mean relation for
the main sequence in the CMD and exclude all stars which deviated by more 
than three sigma from the mean, eliminating both field stars and probable
binaries. In this situation, where the field stars contribute as much, if
not more, to the region of the main sequence, definition of the cluster
main sequence can become an exercise in selection bias.

Fortunately, the availability of multiple indices makes the selection process
much less subjective. To demonstrate the point, we make use of the
filter pair with the largest baseline, $u-y$. For our next cut, we
include stars with $V$ brighter than 17.25 and bluer than
$b-y$ = 0.7, at least 2 observations
each in $y,b,v$, and $u$ and a standard error of the mean in $c_1$ below
0.025 mag. The last constraint guarantees that the error in $u-y$ is below
0.020. The ($V$,$u-y$) CMD is shown in Fig. 4. The increased temperature
sensitivity of $u-y$ is obvious and allows easy identification of
stars that are too blue for their apparent magnitude (field stars),
plotted as open triangles in the figure, and
stars that are too red (binaries or field stars), plotted as crosses.
The success of the process is measurable in Fig. 5, where the 
traditional CMD for the
stars is presented.  With one exception, the stars that lie within the
binary sequence in the standard CMD have been identified as such from the
$u-y$ index. Elimination of the stars that scatter blueward in the $u-y$
figure would remove all the scatter to the blue in the standard CMD.
Though it is probable that elimination of these deviants from the analysis
will sweep away a few true cluster members, the critical point is that
the remaining sample contains a high percentage of single stars whose
structure and evolution are typical of the cluster as a whole.

\section{Fundamental Properties: Reddening and Metallicity}
The select sample of probable members (filled circles) in Fig. 5 includes
99 stars. Our final restriction is to impose a color cut on the turnoff.
Only stars in the color range from $b-y$ = 0.53 to 0.60 will be used
to determine the reddening and metallicity. The blue limit removes the
blue stragglers from the analysis. Since these stars have, by definition,
anomalous evolution, conclusions based upon their photometry are always
open to question. The red cutoff removes the evolved stars of the
subgiant branch and keeps the sample within a color range where the
intrinsic calibrations of the $uvby$ system can be expected to produce
reliable parameters. The final sample includes 71 stars, none of which
have errors in H$\beta$ above $\pm$0.020; only 2 stars have photometric
uncertainty in H$\beta$ above $\pm$0.015.

\subsection{Reddening}
As discussed in Paper I, derivation of the reddening from intermediate-band
photometry is a straightforward, iterative process given reliable estimates
of H$\beta$ for each star. Briefly, the intrinsic $b-y$ color is
derived from the reddening and metallicity-independent H$\beta$ index. Since
$b-y$ is affected by both metallicity and evolutionary state, the intrinsic
value is then adjusted to correct for the differences  between the star
and the unevolved stars used to define the standard relation linking
$b-y$ and H$\beta$. Because the indices estimating the metallicity 
and luminosity are
influenced by the reddening, one must iterate the procedure, correcting each
stellar index for the derived reddening, calculating a new set of intrinsic
indices, luminosities, and metallicities, then re-deriving a revised
reddening. The procedure converges quickly. For cluster stars, all
members are assumed to have the same reddening and metallicity, so one may
calculate the reddening under different fixed assumptions for metallicity, and
the metallicity under different fixed assumptions for the
reddening. Only one combination of reddening and metallicity will produce
mutually consistent calculations. 

Since both procedures give effectively
identical results, the primary decision is the choice of the standard
relation for H$\beta$ versus $b-y$ and the adjustments required to correct
for metallicity and evolutionary state. The two most commonly used relations
are those of \citet{OLS88} and \citet{SN89}. As found in Paper I for
IC 4651, both
produce very similar but not identical results. We have processed the
indices for the 71 stars through both relations and find $E(b-y)$ =
0.192 $\pm$ 0.017 (s.d.) with \citet{OLS88} and $E(b-y)$ = 0.188 $\pm$
0.015 (s.d.) with \citet{SN89}. As a compromise, we will take the weighted
average of the two and use $E(b-y)$ = 0.190 $\pm$ 0.002 (s.e.m.) or
$E(B-V) = 0.260 \pm$ 0.003 (s.e.m.) in the analysis that follows. 

It seems useful to discuss the reliability of the reddening estimate 
before deriving the metallicity
in the next section. Though we have used the turnoff stars alone 
in the calculation,
we can check the consistency of the final value using all the stars in the
members sample, i.e., by including the blue stragglers. Fig. 6 shows a 
plot of the standard relation between H$\beta$ and $b-y$, reddened by 
$E(b-y)$ = 0.190 and superposed upon the observed data for NGC 6253.
With the exception of 3 blue stragglers (indicated by filled symbols)
that lie systematically above the
standard relation, the adjusted curve falls nicely through the mean of the
hotter points defined by the A through early F stars. Note that the curve
does not pass through points defined by the stars at the turnoff. The reason
is that the metallicity effect on the cooler stars is significant and no
metallicity correction has been applied to the cooler end of the standard
relation. The systematic difference relative to an
approximately Hyades metallicity is the first indication of the exceptional
metallicity of the cluster. 

As for the three stragglers located above 
the mean relation, the significantly lower reddening estimate
implied for these stars is explained if they are foreground
stars rather than members of the cluster. This is confirmed
by $c_1$ indices for these stars which are systematically
smaller by approximately 0.1 mag than the expected values at
their $b-y$ based upon the
rest of the stragglers. If we apply the cluster reddening to these stars,
the corrected indices would generate physically implausible combinations of
$(b-y)_0$ and $c_0$. Moreover, use of the individual reddening values
for these stars rather than the adopted cluster value produces metallicity
estimates close to those typical of stars in the solar neighborhood, 
rather than the exceptionally high value found for the cluster.

How does the derived reddening value compare with other estimates?
At the galactic coordinates of the cluster, the reddening map of
\citet{SC98} based upon infrared emission by dust gives $E(B-V)$ = 0.35.
Since the cluster is located more than 150 pc away from the galactic plane
and the reddening map includes all reddening along the line of site
at and beyond the cluster, the difference between the limit and the 
observed value is understandable.

The derived value falls nicely within the range of $E(B-V)$ = 0.23 
to 0.32 found by
BR using comparisons with theoretical cluster CMDs but the 
agreement may be fortuitous if the problems with the broad-band photometry
discussed in the Appendix are tied to the data of BR. The revised values of
$B-V$ from BR transformed to the systems of SA and PI expand the color
differential between the giant branch and the turnoff, independent of 
issues tied to the zero-points, leading to a revision of what age and
metallicity combinations supply the optimum fit to the isochrones. 

Both PI and SA use multicolor CMDs matched to standard sequences (PI)
and theoretical isochrones (SA) to derive $E(B-V)$ = 0.20 and $E(V-I)$
of 0.25 to 0.27. PI quote
uncertainties of $\pm$0.05 mag based upon the internal scatter of the
main sequence alone, while SA provide no error estimate. The fact that
both investigations produce similar estimates of $E(B-V)$ and $E(V-I)$
may be purely fortuitous. PI matches the cluster to
a $BV$ sequence tied to stars typical of the solar neighborhood, i.e.,
metallicity less than or equal to solar, while SA fits isochrones with
Z = 0.05, more that double solar metallicity. The $VI$ diagrams should
be relatively insensitive to metallicity effects, but comparison of
SA and PI demonstrates a zero-point discrepancy in their $(V-I)$ colors 
of 0.14 mag. In summary, while realistic appraisal of the previous cluster 
reddening analyses indicates an estimate of $E(B-V)$ = 0.2 $\pm$0.1,
this result doesn't offer much value in testing the reliability of 
the intermediate-band result.

\subsection{Metallicity from $m_1$} 
                      
Given the reddening of $E(b-y)$ = 0.190, the derivation of [Fe/H] from
the $m_1$ index is a straightforward procedure. The $m_1$ index for a 
star is compared to the standard relation at the same color and the
difference between them, adjusted for possible evolutionary effects, is a
measure of the relative metallicity. Though the comparison of $m_1$ is most
commonly done using $b-y$ as the reference color because it is simpler to
observe, the preferred reference index is H$\beta$ due to its insensitivity
to both reddening and metallicity. Changing the metallicity of a star will
shift its position in the $m_1$ - $(b-y)$ diagram diagonally, while moving
it solely in the vertical direction in $m_1$ - H$\beta$. Moreover, reddening
errors do not lead to correlated errors in both $m_1$ and H$\beta$.

The primary weakness of metallicity determination with intermediate-band filters
is the sensitivity of [Fe/H] to small changes in $m_1$; the typical slope of
the [Fe/H], $\delta m_1$ relation is 11. Even with highly reliable photometry,
e.g., $m_1$ accurate to $\pm$ 0.015 for a faint star, the uncertainty in [Fe/H]
for an individual star is $\pm$0.17 dex from the scatter in $m_1$ alone.
When potential photometric scatter in H$\beta$ and $c_1$ are included,
errors at the level of $\pm$0.2 dex are common. As noted in 
previous papers in this
series, the success of the technique depends upon both high internal accuracy 
and a large enough sample to bring the standard error of the
mean for a cluster down to statistically useful levels, i.e., below 0.10 dex.

Because of the size of the sample, we can minimize the impact of 
individual points such as binaries and/or the
remaining non-members, though they will clearly add to the dispersion.
After correcting each star for the effect of $E(b-y)$ = 0.190, the $m_1$,
$c_1$, and H$\beta$ values for the 71 core cluster members were 
run through the metallicity calibration
using the relation derived by \citet{OLS88} and also applied to the metal-rich
open cluster IC 4651 in Paper I. Before discussing the results, it should be
noted that the range in $H\beta$ among the turnoff sample is small. 
Since all the stars have approximately the same color, variations in 
$m_1$ should be primarily a product of
the evolutionary state, normally measured via $\delta$$c_1$, the difference
between the observed $c_1$ and the standard value at the same color.
As demonstrated in \citet{NT87} from photoelectric observations of M67, for the
cooler F stars, evolution reduces the observed value of $m_1$ at a given color,
making the more evolved stars appear more metal-deficient. Since this effect
occurs exactly within the color range of our sample, we have adjusted the
$m_1$ indices of the more evolved stars upward by 0.3$\Delta$$c_1$, where
$\Delta$$c_1$ is the mean $c_1$ for a star minus the average $c_1$ value 
of the stars fainter than $V$ = 16.5.
We emphasize that while the brighter stars do, in the mean,
have smaller $m_1$ and larger $c_1$, the correction implied by the above
procedure is small and does not completely eliminate the correlation between
$m_1$ and $V$.

From 71 stars, the average $\delta$$m_1$ is --0.071 $\pm$ 0.002 (s.e.m.),
which translates into [Fe/H] = +0.90 $\pm$ 0.03 on a scale where the 
Hyades has [Fe/H] = +0.12. The $\delta$$m_1$ index and its implied 
metallicity are unmatched by any star or cluster observed to date within 
the Milky Way. Though the extreme nature of the estimate casts doubt on its
validity, before evaluating what this means, we turn to another metallicity
indicator, $hk$, as a means of constraining our potential solutions.  

\subsection{Metallicity from $hk$}
The $hk$ index is based upon the addition of the $Ca$ filter to the traditional
Str\"{o}mgren filter set, where the $Ca$ filter is designed to measure the
bandpass which includes the H and K lines of Ca II. The design and development
of the $Caby$ system have been laid out in a series of papers discussing
the primary standards \citep{ATT91}, an extensive catalog of field star
observations \citep{TAT95}, and calibrations for both red giants \citep{ATT98}
and metal-deficient dwarfs \citep{ATT00}. Though the system was optimally
designed to work on metal-poor stars and most of its applications have
focused on these stars \citep{ATC95,BD96}, early indications that the system retained
its metallicity sensitivity for metal-rich F dwarfs have been confirmed
by observation of the Hyades and analysis of nearby field stars \citep{AT02}.

What makes the $hk$ index, defined as $(Ca-b)-(b-y)$, so useful even at the
metal-rich end of the scale is that it has half the sensitivity of $m_1$ to
reddening and approximately twice the sensitivity to metallicity changes,
i.e., $\Delta {\rm[Fe/H]}/ \Delta hk =$5.6. The metallicity calibration for F stars
derived in \citet{AT02} used $\delta hk$ defined relative to $b-y$ as the
temperature index. To minimize the impact of reddening on metallicity, this
calibration has been redone using H$\beta$ as the primary temperature index.
The Hyades $hk$,$(b-y)$ relation has been converted 
to $hk$, H$\beta$ through the standard $(b-y)$, H$\beta$ relation
discussed in Sec. 4.1 and illustrated in Fig. 6.  From 50 stars compiled
in \citet{AT02}, one finds
 
\medskip
\centerline{[Fe/H]$ = -3.51 \delta hk(\beta) + 0.12$}
\smallskip
with a dispersion of only $\pm 0.09$ dex about the mean relation. Though the
derived zero-point of the relation was found to be +0.07, it has been
adjusted to guarantee a Hyades value of +0.12, the same zero-point used for
the $m_1$ calibration. 

We have derived the cluster metallicity using the $hk$ indices for the same
71 stars analyzed above using both $hk$ relative to the $b-y$ relation and
$hk$ relative to H$\beta$. In order to be consistent with our approach to
$m_1$, we have included an evolutionary correction to $hk$ of the same
manner and size adopted for $\delta$$m_1$. The necessity for such a
correction is confirmed by a plot of the mean $hk$ versus $V$ where one
finds that the more evolved stars have systematically lower values of $hk$,
though the size of the trend is comparable to that found for $m_1$. 
The results relative to $(b-y)$ and H$\beta$ are 
[Fe/H] = +0.84 $\pm$ 0.14 (s.d.) and +0.68 $\pm$ 0.14 (s.d.), respectively.
The first point to note is the significantly smaller size of the dispersion
for $hk$-based [Fe/H] in comparison with $m_1$, despite that fact that there
are fewer $Ca$ frames than $v$ frames used in the construction of the indices
and the errors in $hk$ are comparable to or larger than those for $m_1$.
This is a direct benefit of the smaller slopes in the calibration curves.

Second, the [Fe/H] from the $(b-y)$-based relation is larger by a significant
amount, though similar to the $m_1$ abundance. The difference is a reflection
of the high sensitivity of the [Fe/H] to small changes in $b-y$, particularly
through the adopted reddening. If we lower the reddening by 0.02, i.e., increase
the intrinsic $b-y$ by the same amount, the $(b-y)$-based [Fe/H] declines by
0.22 dex. The same reddening adjustment only alters the H$\beta$-based [Fe/H]
by 0.01 dex. Thus, demanding that both approaches give identical results would
require a shift of about --0.015 in $E(b-y)$, not an implausible change given
the photometric uncertainty. A shift of this size only alters the $m_1$-based
[Fe/H] by 0.07 dex, producing an average for $hk$ and $m_1$ of [Fe/H] =
+0.75.

Third, though the analyses of $m_1$ and $hk$ lead to [Fe/H] ranging from +0.65
to +0.9, an annoying range given the photometric accuracy, there is complete
qualitative agreement in the sense that both indices imply that this cluster
is more metal-rich than any cluster or object ever studied on either system.
Assuming that the extreme nature of the indices is not the product of
photometric error, the unique [Fe/H] estimates derived by 
extrapolation of the standard
calibrations brings the validity of their absolute numerical value into
obvious question. In short, NGC 6253 resides in unexplored photometric
territory and the initial foray into this strange land leads to the
conclusion that it cannot be regarded as minor variant on what we are
familiar with locally. Since both photometric systems have been
tested for stars up to [Fe/H] = +0.3, it seems reasonable to assume that the
cluster abundance is, at least, this high. As a simple compromise, we will
assume that [Fe/H] lies somewhere between +0.4 and +0.6. 

Before discussing the age and distance determination, the effects of
errors in the photometry should be investigated. It has already been noted
above that a small shift in the reddening would help to bring the photometric
results into better agreement, though still at a high value of [Fe/H]. Small
errors in $E(b-y)$ can result from independent errors in H$\beta$,
$b-y$ or both, or from a failure of the intrinsic color calibration at high
[Fe/H]. Likewise, one could make the claim that the high metallicity is a 
product of errors in the zero-points for the $m_1$ and $hk$ transformations.

The weakness in most explanations that attempt to resolve the question by
citing an error in an individual index or parameter is that they only
resolve a portion of the problem.  For example, a zero-point error in $m_1$
has little impact on $hk$. Adjusting the reddening due to a claim that the
intrinsic color relation underestimates the metallicity 
effect on $b-y$ would significantly
alter metallicity indices tied to $b-y$, but have little impact on those
based upon H$\beta$. The only exception to this rule would be an error in
the zero-point for the H$\beta$ transformation. If we adjust the H$\beta$
indices by --0.030, the cluster reddening declines by the same amount. The
combined effect on the metallicity is to lower [Fe/H] from $m_1$(H$\beta$),
$hk$(H$\beta)$, and $hk$($b-y$) to +0.47, +0.33, and +0.47, respectively.
The agreement among the indices is better and 
at a somewhat lower metallicity, but it still places
NGC 6253 above the current limit of the metallicity scale. We also
emphasize that every test we have made to check the zero-point of the H$\beta$
photometry, night-to-night comparisons, standard star observations that
bracket the cluster on a given night, indicate that if an offset is
required, it must be small and it must make H$\beta$ larger, not smaller.

\section{Fundamental Parameters: Distance and Age}

\subsection{The Broad-Band CMD}

Given the reddening and metallicity, the traditional approach to deriving
the distance and age is to compare the cluster CMD to a reliably defined
set of theoretical isochrones. By reliably defined we mean color
transformations and bolometric corrections between the theoretical 
and the observational plane that
reproduce the colors and absolute magnitudes of nearby stars with known
temperatures and abundances. From an age standpoint, the critical test
is whether or not a star of solar mass, composition, and age resembles
the sun. For open clusters of solar and subsolar abundance, the impact
of this issue on cluster ages and distances has been emphasized on a
variety of occasions \citep{TAT89,TAM,DA94,TAH,TAB}.

For NGC 6253, two problems immediately arise. First, isochrones 
are invariably created for broad-band systems and a check of the most
recent publications shows that the theoretical isochrones are available
on the $UBVRI$ system, among others, but not $uvby$. Second, though
isochrones have been generated with a mass fraction of metals up to five times
the solar value, the questions of the related helium fraction and whether
or not the metals should be scaled-solar or enhanced in the $\alpha$-elements
remain open. Moreover, the lack of a significant population of such stars
with reliably determined fundamental parameters makes the color transformations
an exercise in faith in the validity of extrapolating synthetic colors from
model atmosphere codes.

Fortunately, the first problem is easily solved. The $b-y$ index has long
been known to correlate well with $B-V$ at a given [Fe/H] with little
dependence on evolutionary state. Though one could use the published
$BV$ data to construct a CMD of cluster members, there are 
zero-point differences
among the broad-band surveys, the photometric accuracy of the surveys is not
uniform, and the surveys only include stars within the core of the cluster.
Instead, we chose to transform our $b-y$ data to $B-V$. The construction of
the mean broad-band photometric indices through the merger of the
published data is discussed in the Appendix; for reasons noted there, we have
chosen the $B-V$ zero-point and scale of SA as the base for the merger. 
Using $B-V$ and $b-y$ for
all the stars brighter than $V$ = 16.5 common to the merged $BV$ sample and
our $by$ data, linear relations were derived between the two indices.   
To optimize the color transformation, the sample was subdivided into three
color regimes, with the transition points at $b-y$ = 0.47 and 0.71. Excluding
4 stars with residuals about the mean relation greater than 0.10 mag, the
transformation curves over the three color regimes are:

\medskip
\centerline{$b-y \leq 0.47$, N=30}
\smallskip
\centerline{$B-V$ = 1.492($\pm$0.038)*$(b-y) + 0.009(\pm$0.016)  }
\medskip
\centerline{$0.47 \leq (b-y) \leq 0.71$, N=264}
\smallskip
\centerline{$B-V$ = 1.717($\pm$0.026)*$(b-y) - 0.096(\pm$0.015)   }
\medskip
\centerline{$(b-y) \geq 0.71$, N=85}
\smallskip
\centerline{$B-V$ = 1.464($\pm$0.029)*$(b-y) + 0.084(\pm$0.024) }
\smallskip

The standard deviations of the residuals about the mean relation 
for the three regions are 0.017, 0.020, and 0.017 mag, respectively, consistent
with a typical uncertainty of 0.015 mag in $B-V$ and less than 0.010 mag in
$b-y$. A plot of the data and the mean relations is shown in Fig. 7.

To define the cluster CMD, we restrict our sample to stars within the core
for $13.0 \leq V \leq 17.25$ with a standard error in the mean for 
$b-y$ below 0.010 (see Fig. 3). The only additional exclusion will be the
stars identified in Fig. 4 by triangles and tagged as nonmembers
based upon a $u-y$ index that places them blueward of the cluster main 
sequence. For $V \leq 13.0$, all stars observed within the entire
field will be used in an effort to strengthen the definition of the red
giant clump and the giant branch. 

\subsection{The Isochrones}
The two primary sets of theoretical isochrones available in an array of 
broad-band systems, metallicities, and ages are those 
of \citet{GI02} (hereinafter referred to as PAD) and \citet{SCH00} (hereinafter 
referred to
as GEN). Based upon the many comparisons,
including our own \citep{ASH,DA94,AHT,TAH}, between open clusters and past and
present generations of isochrones, we will only make use of isochrones that
include convective overshoot mixing. On a scale where solar metallicity is
Z = 0.019 and Y = 0.273, PAD supply isochrones with (Y,Z) = (0.32,0.040) 
and (0.39, 0.070) with either scaled-solar abundances or 
$\alpha$-enhanced abundances. In contrast, the GEN set assumes solar 
(Y,Z) = (0.30,0.02) and has isochrones available for (0.34,0.04) and 
(0.48,0.10). While the latter metallicity is closest to the cluster abundance
implied by the photometry, the combination with the high helium abundance
produced a CMD morphology that was so discrepant in comparison with the cluster
that inclusion of the set added nothing to the discussion. While in the past we
have attempted to test the zero-points of the isochrones by comparison to
the sun and/or field stars \citep{TAB}, there is no way of knowing if a
difference found between theory and observation at solar metallicity transfers
directly to isochrones at [Fe/H] = +0.3 to +0.6. Because we are fixing the
reddening at $E(B-V)$ = 0.26, differences among the derived ages and distances
are solely a product of the differences among the models, both in assumed
composition and in the transformations between the theoretical and the
observational plane.

For the first isochrone fit, we use the GEN models for (Y,Z) = (0.34, 0.04),
equivalent to [Fe/H] = +0.3, the low end of the metallicity range
potentially applicable to NGC 6253. The isochrones in Fig. 8 have been
shifted by +0.26 in $(B-V)$ and by 11.60 in $V$, the apparent distance
modulus. Filled circles and crosses are the same core members and probable
binaries identified in Fig. 4. Open triangles are the additional data from
the core of the cluster outside the color range of Fig. 4 and squares are
the points outside the core brighter than $V$ = 13.0. The superposed 
isochrones have ages of 3.2, 3.55, and 4.0 Gyrs. The match to the isochrones
at the turnoff and the giant branch is very good, though the
curvature of the main sequence turnoff is slightly larger than predicted
by the models. Because of the inclusion of the brighter stars over the
entire field, the giant branch can now be mapped to the models well above
the level of the clump at $V$ = 12.7. The overall fit of the data 
implies an age between 3.4 and 4.0 Gyr, so we will adopt 3.7 $\pm$ 0.3
as the value for these models.

The next set of models will be those of PAD with scaled-solar abundances
and (Y,Z) = (0.32, 0.04) or [Fe/H] = +0.32. Despite the similarities in
[Fe/H] to the isochrones of Fig. 8, a match to the turnoff and main 
sequence requires an apparent modulus of 12.30 in Fig. 9. The ages
of the superposed isochrones are 2.0, 2.5, and 3.2 Gyrs. The overall
quality of the fit is significantly reduced. The turnoff exhibits the
same difference in curvature found in Fig. 8, but the subgiant branch
lies between the two older tracks while the color of the turnoff
is between the two younger. More important, the color of the isochrone
giant branches is too red by about 0.2 mag in $B-V$ and the red giant
clump is too faint compared to the observations by 0.6 mag. If one
reduces the reddening to $E(B-V)$ = 0.20, the apparent modulus is
lowered to 11.95, producing a consistent match between the subgiant
branch and the turnoff color, but leaving the clump too faint by 0.25
mag and the giant branch of the isochrones too red by about 0.08 mag.

The next alternative is the PAD models with (Y,Z) = (0.32,0.04) but
enhanced-$\alpha$ elements. The results are shown in Fig. 10 for ages of
2.0, 2.5, and 3.2 Gyrs and an apparent modulus of 12.05. In contrast
with the previous isochrones, the match from the main sequence through
the base of the giant branch is effectively perfect in magnitude and color.
The curvature of the turnoff is exactly as predicted. The giant branch of
the isochrones is still too red by between 0.03 and 0.06 mag, but the
clump is much closer in luminosity to the observed data. The age of
the cluster is easily constrained to be 2.5 $\pm$ 0.1 Gyr. 

The last match shown in Fig. 11 is that for the PAD models, $\alpha$-enhanced, 
with 
(Y,Z) = (0.39,0.07) or [Fe/H] = +0.57. The adopted apparent modulus 
is 12.15 and the
isochrones have ages of 1.0, 1.4, and 2.0 Gyr. As in Fig. 9, the quality of
the fit is disappointing, particularly since this represents an overall
metallicity closest to that defined by the photometric indices. 
The observations of the turnoff and subgiant
branch cannot be reconciled with a single age and the discrepancy between the
locations of the giant branch is large, with the usual pattern of the models
being too red and the theoretical clump too faint. As with Fig. 9 one could
slightly improve the turnoff match by lowering the reddening and shifting an
older isochrone into alignment with the cluster, but the subgiant branch
would still be found at a magnitude level inconsistent with the age defined
by the color of the turnoff. Use of the scaled-solar isochrones at Z =
0.07 produces an increase in the apparent modulus to 12.55 and an even
worse match to the shape of the turnoff and the position of the giant
branch. 

Given the large range of parameters derived using the various isochrones,
can we reach any plausible conclusions regarding the 
cluster age and distance? The first point to note is that, with the 
exception of the Z = 0.07 isochrones, the models used were of lower Z
than implied by strict acceptance of the photometric metallicities.
Increasing Z for a cluster at a fixed reddening leads to a younger age
and a larger distance modulus, as confirmed above. If the age and reddening
were known, one could derive the metallicity by selecting the isochrones
that optimally matched the CMD, a simplified version of the approach
taken by BR.

As discussed earlier, the morphology of the CMD, in particular the existence
of a main sequence hook and gap, constrains the cluster to be younger than
about 5 Gyr. A more quantitative answer can be attained by using the
morphological age ratio (MAR) as discussed and revised by \citet{ATT85} and
\citet{TAT89}. Based upon the ratio of the difference in $V$ between
the red giant clump and the brightest point at the turnoff and the 
difference in $B-V$ between the giant branch at the level of the clump
and the vertical turnoff, it has the advantages of being more sensitive
to age changes than either difference alone, reddening-independent, and
less sensitive to metallicity than either difference alone. From the
CMD of NGC 6253, the MAR value is found to be 2.9 (2.8 if one adjusts for
the possible difference in reddening for the giants compared to the
turnoff stars \citep{TAT97}), slightly smaller than M67
at 3.25 but larger than NGC 6819 at approximately 2.0. If we adopt ages
of 4.0 Gyr \citep{DIN} and 2.5 Gyr \citep{RV98} for these clusters, the
approximate age of the NGC 6253 is 3.5 Gyr, in very good agreement with
the values typically derived by BR from a more involved match to a variety
of isochrones. Keeping in mind the caveat that we have no knowledge of how
reliable an index such as MAR is for [Fe/H] of +0.4 or higher, the GEN
isochrones are consistent with this age for the cluster if [Fe/H] is
increased to +0.4. At this metallicity, the apparent modulus becomes 11.7,
leading to $M_V$ for the clump of +1.0, consistent with expectation
for an older cluster of high metallicity as empirically determined by
\citet{TAT97}.

A unfortunate issue that we cannot resolve is the disagreement between
the scaled-solar isochrones of GEN and PAD. Though the GEN models use
Y higher by 0.02, the differences in age ($\sim$ 1 Gyr) and apparent
modulus (0.7 mag) seem excessive given the similar values of Z. The
lower age defined by the turnoff color is consistent with the 
discrepancy between the colors of the observed and computed giant
branches; the observed CMD morphology requires an older age than 
allowed by the absolute colors. This explains why artificially lowering the
reddening produces a better match to an older isochrone and why increasing
Z to 0.07 makes the problem worse. Following the pattern, the predicted
$M_V$ values of the clump of +0.4 and +0.15 for solar-scaled Z = 0.04 and 0.07,
respectively, are inconsistent with a very metal-rich cluster at these ages.

Despite the problems, the PAD isochrones may provide a solution to the
entire issue. By far, the best match to the shape of the observed
turnoff and subgiant branch was supplied by the $\alpha$-enhanced isochrones
with Z = 0.04. The improvement over any solar-scaled set may indicate that
the true [Fe/H] for NGC 6253 is closer to +0.4, but the photometric
indices, $m_1$ and $hk$, give anomalous abundances because the 
energy distributions of
the stars are distorted by the enhanced $\alpha$ elements. The morphology
of the CMD and the $M_V$ of the clump cannot be compared reliably to
nearby clusters of the same age because these clusters have predominantly
scaled-solar distributions. The $M_V$ of the clump with the optimal match
of Fig. 10 is +0.65, within the observed range for nearby clusters of the
same age and close to the value predicted by the isochrones.

\section{Summary and Conclusions}                                     

The open cluster, NGC 6253, has unusual significance in two ways. It
populates a rather rare class of open clusters, old and significantly closer
to the center of the Galaxy than the sun. The statistical studies of
\citet{JP94} and \citet{HUF} have demonstrated that, observational 
difficulties aside,
the number of open clusters observed within 7 kpc of the galactic
center is so small that the probability for survival beyond 1 Gyr is
low; NGC 6253 falls just outside this zone. Second, only the old open
cluster NGC 6791 has been identified as having a metallicity significantly
higher than the Hyades \citep{CHA}, and even this claim has been questioned
\citep{TAY}. Past attempts to quantify the key parameters for NGC 6253
have depended primarily on consistency fits to multicolor, broad-band
relations and synthetic CMD's tied to theoretical isochrones. Though
there is a consensus linked to the CMD morphology that the cluster must be
2 to 4 Gyr old and more metal-rich than the sun, narrowing the range
has been difficult. Though the source
of the disagreement has originated mostly with the different predictions
from the various isochrones, a problem repeated in Sec. 5, the limitations
imposed by the differences among the photometric surveys must be accounted
for. As noted in the Appendix, the absolute answer one derives for
the parameters would depend upon which zero-point and which color slope
one chooses as the $correct$ system for defining the broad-band CMD and
color-color plots.

We reiterate one of the chief advantages of the intermediate-band approach 
because it allows to fix the reddening through H$\beta$ with
only a weak dependence on the cluster metallicity. Removing this as a
free parameter in any match to theoretical isochrones significantly
restricts the allowed combinations of metallicity, distance, and age.
As has now been demonstrated using IC 4651 (Paper I), NGC 6397 (Paper II), 
and NGC 6253, it is possible to derive the distance with an internal
error of only $\pm 0.1$ mag, with internal errors of the mean for
$E(b-y)$ and [Fe/H] of $\pm 0.002$ and $\pm 0.02$, even though we cannot
guarantee that all non-members and binaries have been eliminated.

Given this precision, how does one explain the extraordinary 
metallicities implied by both the $m_1$ and $hk$ indices?
The flip side to the internal precision that is achievable is that 
the primary source of uncertainty in deriving
photometric-based parameters from a large statistical sample falls upon
the transformations of the instrumental indices to the standard system.
The challenge of doing this well, even for an established
broad-band system like UBV, is obvious from the comparisons of the 
data on NGC 6253 as given in the Appendix. However, as noted in Sec. 4,
to simultaneously reduce the cluster metallicity by claiming an 
error in the photometry requires either a conspiracy between $m_1$ and
$hk$ or an error in the common link for the reddening and the both 
metallicity indicators,
H$\beta$. Unfortunately, an error of over 0.03 mag would be required to
have a significant impact on the metallicity and, even for this change,
would produce [Fe/H] well above the Hyades.

A probable solution to the question is that the indices are, in fact,
close to the truth but their interpretation is based upon a questionable
extrapolation of the standard relations for stars with more "normal"
abundances, both in an absolute and a relative sense. This distortion could
apply to both the metallicity and the reddening determination, though
the size and direction of the effect on the latter variable are unknown.
The existence of $uvby$ index distortions due to abundance anomalies has been 
known for over 20 years with CN variations in CH and Ba stars producing
anticorrelated changes in $m_1$ and $c_1$ being a prime 
example \citep[see, e.g.,][]{BO80,LB82}. The correlated 
impact of abundance anomalies
on $m_1$, $c_1$, and $hk$ has been documented for field stars \citep{ATT91,AST}
and globular clusters like M22 \citep{ATC95} and $\omega$ Cen \citep{RE00}, but
the objects under discussion have been metal-poor giants contained within
systems that exhibit star to star variations, either due to inhomogeneities
at the time of formation or variations in the evolutionary impact of 
post-main-sequence internal adjustments. If the source of the anomalous
indices in NGC 6253 is an enhanced distribution of $\alpha$-elements, it is
common to all the stars in the cluster and must be the product of galactic
chemical evolution. To our knowledge, NGC 6253 is the first open cluster to
exhibit potential evidence for a non-scaled-solar element distribution.

How does this solution mesh with observations of the Galactic disk interior
to the sun? Two observational pieces of evidence lend some support to this
option, though the significance of each remains indeterminate either because
the result is too recent to be considered established or because the conclusions
remain controversial. One piece of corroborative evidence is supplied by 
spectroscopic analysis of Cepheids interior
to the solar circle by \citet{AN12}, indicating that Cepheids that lie more
than 1.5 kpc closer to the galactic center than the sun have [Fe/H] between
+0.2 and +0.4, significantly higher than those found in the solar neighborhood.
From simple chemical evolution models, the dramatically higher [Fe/H] should
be typical of recently formed stars within this galactocentric radius.
It must be emphasized that though the error bars on [Fe/H] are small, so
is the sample. If this inner discontinuity is real, it needs to be confirmed with
a larger sample extending closer to the galactic center and an explanation 
for its apparent absence from HII region analyses \citep{HW99} would be useful.

Another piece of the puzzle is supplied by another well-studied, older open 
cluster located at a galactocentric distance
comparable to NGC 6253 --  NGC 6791, with a controversial history
of reddening, metallicity, and age estimation. The most recent comprehensive
analysis of the cluster is given by \citet{CHA}, where CMD comparisons to
theoretical isochrones restrict the cluster metallicity to [Fe/H] = +0.4 $\pm$
0.1, in excellent agreement with the sole high dispersion spectroscopic 
abundance analysis of a horizontal branch star by \citet{PG98}. (For an alternative
view of this discussion, see \citet{TAY}.) If the single spectroscopic result can
be trusted as typical of the cluster, it is intriguing to note that the star
shows enhancements relative to solar ratios of Mg, Si, Na, and N, but not Ca.
Even accepting the high and enhanced metallicity for NGC 6791, its
unique properties within the cluster sample, combined with its highly
eccentric orbit \citep{SC95},
have led some to question where it fits within the population mix of the
Galaxy, thin disk, thick disk, or bulge. If the last alternative is correct, the
cluster is not relevant to understanding the evolution of the disk. If our
conjecture regarding the metallicity distribution of NGC 6253 is correct,
a trend of high metallicity beginning at least 8 Gyr ago and extending to the
present appears established for the disk interior to $\sim$7 kpc from the
galactic center, keeping in mind that the current location for the clusters may
not be representative of where they formed.

\acknowledgements
The progress in this project would not have been possible without the time
made available by the TAC and the invariably excellent support
provided by the staff at CTIO. NDL gratefully acknowledges the summer
support provided by the Tombaugh Fund at the University
of Kansas. The cluster basic parameters project has been
helped by support supplied through the General Research Fund of
the University of Kansas and from the Department of Physics and
Astronomy. 

\appendix
\section{Broad-Band Photometric Merger}

Broad-band CCD data are available from the three photometric surveys 
with $UBVRI$ included in BR and SA, and $BVI$ 
found in PI. SA supply a summary of the differences
in the indices among the various studies, finding significant offsets for
a number of the colors. The risks in the standard approach of averaging the
residuals for all the stars in such comparisons are:

(a) The residuals may be color-dependent and will not show significant
variation with $V$ if the color range is large at all $V$, though the
dispersion will be increased;

(b) The number of stars and the errors invariably increase with increasing
$V$. Real trends among the residuals may remain hidden if they exist at a
level comparable to the errors among the faintest stars.

The first indications of a potential problem in the broad-band data
arose during our attempts to compare our $V$ data, tied to the $y$ filter,
with the broad-band sources. Using all the stars in common, SA
noted systematic differences in $V$ in the sense: 
(SA - BR) = --0.13 $\pm$ 0.10 for 437 stars between $V$
= 11 and 21 and (SA - PI) = --0.07 $\pm$ 0.05 for 115 stars between $V$ =
11 and 18. The residuals appear reasonably 
consistent with the 
errors quoted by the investigations since the samples tend to be dominated
by the fainter range of the $V$ distributions. 
Similar comparisons were made between our $V$ data and the above 
investigations and it became apparent that in the case of BR,
while the internal errors were the best of the three studies, a color
dependence existed among the residuals in $V$.

To avoid potential confusion
caused by the inclusion of fainter stars with larger errors, the color
dependence for each survey was tested using only stars brighter than
$V$ = 16. Table 2 summarizes the statistics for the comparisons, including
the number of stars in each sample, the number of stars excluded for deviating 
by more than 0.1 mag from the mean of the residuals, the color relation
derived in the sense (Table 1 - study) $= [a \times (B-V)] + b$, the correlation
coefficient, R, the dispersion
in the residuals for $V$ brighter than 16 after correcting for offsets
and color terms, and the residuals after correction for stars in common
at all magnitudes. The residuals for the transformed 
$V$ as a function of $V$ for all three
studies are shown in Fig. 12. 

The internal errors for BR are excellent, with an increase of only
about one-third in the dispersion of the residuals in going from the brighter
stars ($\sigma = 0.016$) to the entire sample ($\sigma = 0.021$). In contrast,
both SA and PI exhibit a 50 \% growth between the
bright stars and the entire sample and a larger dispersion in both cases
than found for BR. It should be noted that the offset and
dispersion in $V$ found by SA in their comparison with PI
are in excellent agreement with the relative values derived from 
the data in Table 2. Unfortunately, no two photometric studies to date
produce $V$ consistent with each other at the 0.01 mag level, though our
system lies near the mean of the broad-band $V$ data.

Though no simple comparison is available between the remaining $uvby$
colors and the broad-band filters, the broad-band indices can be of use
in refining the cluster parameters, particularly since theoretical isochrones
are generally available on such systems. To optimize the effectiveness of
such analyses, a decision was made to merge all the data from the published
surveys. Because the photometry in all cases has been calibrated by 
transforming each magnitude from an instrumental system to the standard
system using an offset and a color term prior to calculating the
color indices, the published color indices were returned to individual
magnitudes and the differences between the magnitudes in each filter
calculated. The residuals were then checked for an offset and any evidence
of a color term. Because it had the largest overlap with the two remaining
studies, the photometry of BR was used as the reference source,
though it was not adopted as the standard system. The reason is that
the color term found for $V$ carries over to an even greater degree in $B$,
as summarized in Table 2. The size of the color term for $B$ is large, but
identical from the comparisons with SA and PI. Moreover,
the size of the term is comparable to the color term derived by BR
in transforming the instrumental $B$ magnitudes to the standard system, a 
color term that is dramatically larger for $B$ than any other filter, 
including $U$.

For the standard colors, the photometry of SA has been chosen by
applying the same zero-point shift as used in transforming $V$ to our system
to all the broad-band filters. Color terms were only included in transforming
the $V, B, U,$ and $R$ data of BR to the standard system. In all
other cases, the color term was set to zero and a simple offset rederived
from the data used to generate Table 2. The magnitudes for each filter were
then merged, giving the photometry of BR twice the weight of
the photometry of SA and PI individually.

\clearpage
\figcaption[TDLA.fig1.eps]{Standard errors of the mean for the $V$, $b-y$,
$m_1$, $c_1$, $hk$ and H$\beta$ indices as a function of $V$. Major tick 
marks on the $y$-axis are 0.2 mag apart.  The vertical scale has been offset by
0.2 mag for each index for visual clarity. \label{fig1}} 

\figcaption[TDLA.fig2.eps]{Color-magnitude diagram for stars with at least
2 observations each in $b$ and $y$. Filled circles are stars with
an indication of variability. \label{fig2}}

\figcaption[TDLA.fig3.eps]{CMD for stars within $100"$ of the cluster 
center. Crosses are stars with errors in $b-y$ greater than 0.010 mag. 
\label{fig4}}  

\figcaption[TDLA.fig4.eps]{CMD of stars in cluster core using $u-y$ as
the color index. Triangles are stars tagged as nonmembers blueward 
of the cluster mean relation while crosses identify binaries and 
nonmembers redward of the mean relation. \label{fig5}}

\figcaption[TDLA.fig5.eps]{Traditional CMD for the stars in Fig. 4.
Symbols have the same meaning. \label{fig6}}

\figcaption[TDLA.fig6.eps]{H$\beta$ versus $b-y$ for highly probable
cluster members at and blueward of the turnoff; the three filed circles
denote the positions of stars believed to be foreground stars. 
Solid curve is the
standard relation reddened by $E(b-y)$ = 0.190. \label{fig7}}

\figcaption[TDLA.fig7.eps]{Relation between $B-V$ from the merged photometry and
$b-y$ for stars brighter than $V$ = 16.5. \label{fig8}} 

\figcaption[TDLA.fig8.eps]{Comparison of the cluster CMD to the overshoot
GEN isochrones for (Y,Z) = (0.34, 0.04).  The reddening and apparent
distance modulus are $E(B-V) = 0.26$ and $(m-M) = 11.6$.
Filled circles are probable core members, crosses are probable binaries,
triangles are the remaining stars within the core and squares are all stars
brighter than $V$ = 13.\label{fig9}} 

\figcaption[TDLA.fig9.eps]{Same as Fig. 8 for the scaled-solar isochrones
of PAD with (Y,Z) = (0.32, 0.04). \label{fig10}} 

\figcaption[TDLA.fig10.eps]{Same as Fig. 8 for the enhanced-$\alpha$ isochrones
of PAD with (Y,Z) = (0.32, 0.04). \label{fig11}} 

\figcaption[TDLA.fig11.eps]{Same as Fig. 8 for the enhanced-$\alpha$ isochrones
of PAD with (Y,Z) = (0.39, 0.07). label{fig12}} 

\figcaption[TDLA.fig12.eps]{Residuals in $V$ between Table 1 and the cited
reference after adjusting the published data for an offset and a 
color term, if necessary. \label{fig13}} 
\enddocument